\documentclass{amsart}
\usepackage{latexsym}
\usepackage{amsfonts}
\usepackage{graphicx}
\usepackage{epsfig}
\usepackage{amsmath}
\usepackage{amssymb}
\usepackage{mathrsfs}
\usepackage{color}

\begin{document}
\bibliographystyle{abbrv} 
\newtheorem{thm}{Theorem}[section]
\newtheorem{lem}[thm]{Lemma}
\newtheorem{cor}[thm]{Corollary}
\newtheorem{df}[thm]{Definition}
\newtheorem{rem}[thm]{Remark}

\def\proof{{\noindent{\sc Proof.\ }}}
\newcommand{\QED}{\hspace{1ex}\hfill$\Box$\vspace{2ex}}

\newcommand{\eps}{\varepsilon}
\newcommand{\al}{\alpha}
\def\t{\mathbb{\vartheta}}
\def\Ro{\mathbb{R}}
\newcommand{\fm}{f}
\newcommand{\vm}{v^{\tiny{\rm min}}}
\newcommand{\ysm}{y^*_{\tiny{\rm min}}}
\newcommand{\ba}{{\rm ba}(\bar\Omega;\mathbb{R}^3)}
\newcommand{\B}{B(\bar\Omega;\mathbb{R}^3)}
\newcommand{\Ka}{\mathscr{K}_a}

\def\mint{{-}\hspace{-2.5ex}\int}
\def\minint{\mathop{\int\hspace{-2.1ex}{\vspace{-0.5ex}-}}}

\newcommand\LL{\hbox to 7pt{\vrule width.4pt \vbox to 7pt{\vfill \hrule width 5pt height.4pt}}}

\newcommand{\rig}{\mbox{$\mathfrak{R}$}}
\newcommand\transp{^\mathsf{T}}
\newcommand\Haus{\mathcal{H}}
\newcommand\Leb{\mathcal{L}}
\newcommand{\Rr}{\textrm{{I\!\!R}}}
\makeatletter
\newdimen\rh@wd
\newdimen\rh@hta
\newdimen\rh@htb
\newbox\rh@box
\def\rh@measure#1{\setbox\rh@box=\hbox{$#1$}\rh@wd=\wd\rh@box \rh@hta=\ht\rh@box}

\def\widecheck#1{\rh@measure{#1}%
 Ê\setbox\rh@box=\hbox{$\widehat{\vrule height \rh@hta width\z@ \kern\rh@wd}$}%
 Ê\rh@htb=\ht\rh@box \advance\rh@htb\rh@hta \advance\rh@htb\p@
 Ê\ooalign{$\vrule height \ht\rh@box width\z@ #1$\cr
 Ê Ê Ê Ê Ê \raise\rh@htb\hbox{\scalebox{1}[-1]{\box\rh@box}}\cr}}
\makeatother
\def\red #1{{\color[rgb]{1, 0,0} #1}}
\def\green #1{{\color[rgb]{0,1,0} #1}}
\def\blue #1{{\color[rgb]{0,0,1} #1}}
\def\mag #1{{\color[rgb]{1,0,1} #1}}
\def\magenta #1{{\color[rgb]{1,0,1} #1}}

\title{On variational dimension reduction in structure mechanics}

\author[Roberto Paroni]{Roberto Paroni}
\address{
Dipartimento di Architettura Design e Urbanistica,
Universit\`{a} di Sassari} \email{paroni@uniss.it}

\author[Paolo Podio-Guidugli]{Paolo Podio-Guidugli}
\address{Dipartimento di Ingegneria Civile, Universit\`a degli Studi di Roma TorVergata} \email{ppg@uniroma2.it}

\maketitle

\begin{abstract}
The classical low-dimensional models of thin structures are based on certain a priori assumptions on the three-dimensional deformation and/or stress fields, diverse in nature but all motivated by the smallness of certain dimensions with respect to others.
In recent years, a considerable amount of work has been done in order to rigorously justify these a priori assumptions; in particular, several techniques have been introduced to make dimension reduction rigorous. 
We here review, and to some extent reformulate, the main ideas common to these techniques, using some explicit dimension-reduction problems to exemplify the points we want to make.

\end{abstract}
\section{Introduction}

The classical models of thin elastic structures  have big names attached, like Euler, D. Bernoulli, Navier, Kirchhoff and Love, as well as, more recently, Timoshenko, E. Reissner and Mindlin; they are all based on certain a priori assumptions on the three-dimensional deformation and/or stress fields, diverse in nature but invariably motivated by the smallness of certain dimensions with respect to others; they are all low-dimensional, and all admit a variational formulation.
In the past couple of decades, several methods of \emph{variational convergence} have been introduced and used to rigorously justify those classical models. The aim of the present
paper is to review and extend the way variational convergence techniques are used to achieve dimension reduction. We believe that the ideas we are going to present apply also to methods of homogenization \cite{Ta09}, of discrete-to-continuum passage \cite{BDMG99, BrGe04}, and of singular perturbation \cite{MoMo77}, but we shall not discuss those applications here, the cursory remarks in Section \ref{3.4} apart.

The instances of variational convergence we have in mind are
asymptotic expansions, functional analysis methods and, of course, ($G$-, $H$-, and) $\Gamma$-convergence \cite{Sp68, DGSp73, MuTa77, DGFr75}.
We shall not enter in the details of any of these techniques. Roughly speaking, their common and essential trait is that, with their use, \emph{problem convergence implies solution convergence}. 
To put it simply, given a problem sequence $\{P_\varepsilon\}$ and the associated solution sequence $\{u_\varepsilon\}$, 
the variational convergence of $\{P_\eps\}$ to a \emph{limit problem} $P_0$ implies the convergence of $\{u_\varepsilon\}$ to a solution $u_0$ of $P_0$:
\begin{equation}\label{convar}
P_\eps\to P_0\;\Rightarrow\; u_\eps\to u_0.
\end{equation}
Both the physical meaning of solutions $u_\eps$ and the type of their convergence to $u_0$ are essentially determined by the problem sequence. These issues are of no importance to our present discussion. Instead, a relevant issue for us is whether and how the limit problem $P_0$ and its solution $u_0$ are related to a \emph{real problem} $P^r$ and its solution $u^r$.

By a real problem we mean a three-dimensional problem of interest in applications. 
We envisage two situations, the latter occurring more often than the former: 
\vskip 2pt
\noindent (i) given a problem sequence $\{P_\varepsilon\}$ variationally convergent to problem  $P_0$, at least one real problem $P^r$  in reasonably tight kinship with $P_0$ is looked for; 

\noindent (ii) given a real problem $P^r$, at least one variationally convergent sequence $\{P_\eps\}$ is looked for, such that its limit problem $P_0$ is in reasonably tight kinship with  $P^r$. 

It may also happen that in an application community one real problem $P^r$ has been associated with  one or more \emph{approximate problems} $P^a$, whose forms were guessed on the basis of  shrewd combinations of physical intuition and mathematical technique.  This has been the case with problems, real and approximate, coming from the mechanics of thin elastic structures, where a $P^a$ has usually to do with a \emph{low-dimensional} model of the structural body considered in problem $P^r$. For one example among many, think of the real, and hence three-dimensional, plate-like bodies considered in the engineering mechanics community and of their approximate two-dimensional models associated with the names of Kirchhoff-Love and Reissner-Mindlin: their derivations do not pass a rigorous scrutiny, but their predictions are nevertheless  sufficiently accurate for most of the technical purposes. Parallel to efforts to straighten those derivations, attempts have been made  to validate the choice of those approximate problems, and others, variationally, that is, by showing that a given successful $P^a$ coincides with the limit problem $P_0$ of an appropriately chosen problem sequence $\{P_\varepsilon\}$, or because it does not differ much from it. There is then reason to consider a variant of situation (ii), when,
\vskip 2pt
\noindent(iii) given a $P^r$ and an associated \emph{low-dimensional} $P^a$, a variationally convergent sequence  $\{P_\eps\}$ of three-dimensional problems is looked for, such that its limit problem $P_0$ is in reasonably tight kinship with $P^a$. The dimension of $P_0$ may be three or the same as $P^a$'s, depending on the  procedure adopted to carry out the variational limit (see the discussion in Section \ref{sec_cs}); in the latter case, $P_0$ should coincide with $P^a$, in the former it should be possible to rewrite $P_0$ as a low-dimension problem, reducible to, if not coinciding with, $P^a$.

Historically, this third situation has been the first to be explored with the use of variational convergence; $P_0$ is low-dimensional in \cite{ABP91, ABP94}, and is three-dimensional in \cite{BCGR92, Ci97, CiDe79,FJM02, LDRa95}. Unfortunately, the success obtained by employing a certain `natural' choice of problem sequence to achieve dimension reduction by $\Gamma$-convergence  had two undesirable consequences: for one, nobody ever parted with that type of problem sequence; for two, those model problems that could not be validated by the use of a sequence of that very same type where regarded as somehow suspicious, in spite of the indications to change coming from non-variational validation methods, such as the \emph{method of internal constraints} introduced  in \cite{PGJEla} and its development, the \emph{scaling method} of \cite{MPG} (see also \cite{PGCism}). In fact, in \cite{PPGT06, PPGT07, PePG09} those indications have been shown to lead to two non-conventional and different  variational validations of the Reissner-Mindlin plate model.

In this paper, we propose to liberate $\Gamma-$convergence practitioners from the commitment to a standard problem sequence.  In fact, we stress the discovery power intrinsic to nonstandard choices, leading to a variety of limit problems, with their related approximate and real problems, some old and some new.

\section{The standard problem sequence}\label{sec_cs}

To fix ideas and illustrate some of the concepts introduced so far, we examine an explicit problem in structural engineering. Our twofold intention is to exemplify the points we want to make and to summarize the procedure  adopted in most
of the literature to define the sequence of problems $\{P_\varepsilon\}$.

\subsection{Problem $P^r$.}\label{2.1} 
\begin{figure} [h]
\centering 
\def\svgwidth{300pt}
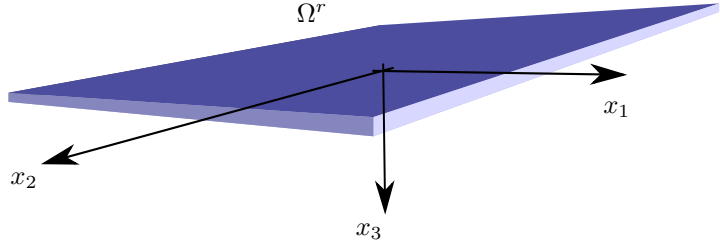
\caption{The domain $\Omega^r$}\label{fig1}
\end{figure}

Consider a plate-like body of thickness $2h^r=2\, cm$ with square cross-section of side length $2\ell^r=200\, cm$. With reference to the Cartesian frame shown in Fig. \ref{fig1}, we denote by
$
\omega^r=(-\ell^r,+\ell^r)\times(-\ell^r,+\ell^r)
$  
the mid cross-section, and identify the body point-wise with the region
$
\Omega^r=\omega^r\times(-h^r,+h^r)
$
it occupies in the reference configuration shown in Fig. \ref{fig1}.
We let $\Omega^r$ be clamped on the Dirichlet part $\partial_D\Omega^r = \partial_D\omega^r\times(-h^r,+h^r)$ of its boundary, subject to null contact loads on the complementary Neumann part, the only applied loads being a distance-force field $b^r$  over $\Omega^r$. Moreover, we assume that $\Omega^r$ is comprised of a linearly elastic material, with elasticity tensor 
$\mathbb{C}^r$.

With a view to finding the equilibrium displacement field $v$ in $\Omega^r$, we let $E(v)$ denote the symmetric part of the gradient of $v$, and we formulate the real problem $P^r$ as follows: 
\vskip 6pt
\noindent minimize the \emph{total-energy functional} 
\begin{equation}\label{realf}
\mathcal{F}^r(v):=\int_{\Omega^r}\Big(\frac 12\mathbb{C}^r [E(v)]\cdot E(v)- b^{r}\cdot v\Big) dx
\end{equation}
over the space $$H^1_D(\Omega^r;\mathbb{R}^3):=\{v\in H^1(\Omega^r;\mathbb{R}^3):v=0 \mbox{ on }\partial_D\Omega^r\};$$ in short,
\begin{equation}\label{Pr_var}
\textrm{find}\;\;u^r=\,\arg\!\!\!\!\!\!\!\!\!\!\!\min_{v\in H^1_D(\Omega^r;\mathbb{R}^3)}\mathcal{F}^r(v)\,.
\end{equation}
Note that the minimizer $u^r$ of problem $P^r$ can be equivalently determined by solving the associated Euler-Lagrange problem:
\begin{equation}\label{Pr_weak}
\begin{aligned}
&\qquad\qquad\textrm{find}\;\,u^r\in H^1_D(\Omega^r;\mathbb{R}^3)\;\, \textrm{such that}\\
&\qquad \int_{\Omega^r}\mathbb{C}^r [E(u^r)]\cdot E(v)\, dx= \int_{\Omega^{r}} b^{r}\cdot v\, dx,
\; \forall v\in H^1_D(\Omega^{r};\mathbb{R}^3)
\end{aligned}
\end{equation}
(uniqueness follows from well-know assumptions of physical plausibility on $\mathbb{C}^r$). 

\subsection{Problem $P^a$} Let us denote by 
$
\eps^r:={h^r}/\ell^r=0.01
$
the \emph{thickness parameter}, that is, the thickness-to-side-length ratio of $\Omega^r$. When $\eps^r\ll 1$, as is the case for the problem at hand, it is quite common in engineering applications to replace problem $P^r$ with an approximate problem $P^a$ posed over the two-dimensional region $\omega^r$.
Oftentimes, when the material $\Omega^r$ is comprised of is isotropic and the load perpendicular to the cross-section plane,
$P^a$ is taken to be the Kirchhoff-Love plate problem. Here are formulations of this two-dimensional problem that parallel, respectively, \eqref{Pr_var} and \eqref{Pr_weak}; for simplicity, we restrict attention to the case when all of the  lateral boundary of $\Omega^r$ is clamped, and hence $\partial_D\omega^r=\partial\omega^r$.

Let the total energy functional of interest be defined over the space
\[
H^2_0(\omega^r;\mathbb{R}):=\{w\in H^2(\omega^r;\mathbb{R}):w=0{\mbox{ and } w_{,n}=0} \,\mbox{ on }\,\partial\omega^r\}
\]
and have the following form:
\[
\mathcal{F}^a(w):=\int_{\omega^r} \Big(\frac{1}{2}\Big( {\bar D}^a(\Delta w)^2-
 {\bar d}^a\big(w_{,11} w_{,22}-(w_{,12})^2\Big)- {\bar b}^{a}w\Big)
 dx,
\]
where, for  $D^a,d^a$ two given positive material constants, 
$
{\bar D}^a={D}^a(h^r)^3$ and ${\bar d}^a={d}^a(h^r)^3
$,
and where 
\[
{\bar b}^{a}:=\int_{-h^r}^{+h^r}b^r(x_1,x_2,x_3)dx_3. 
\]
This functional is stationary if
\begin{equation}\label{KL_var}
\int_{\omega^r} {\bar D}^a \Delta w^a\,\Delta w\,dx
 =\int_{\omega^r}{\bar b}^{a}w\,
 dx,\quad \forall \; w\in H^2_0(\omega^r;\mathbb{R}).
\end{equation}
One seeks to find the unique
$\;\displaystyle{
w^a=\arg\!\!\!\!\!\!\!\!\!\min_{w\in H^2_0(\omega^r;\mathbb{R})}\mathcal{F}^a(w)}
\,$
or, alternatively, to find the unique  $w^a\in H^2_0(\omega^r;\mathbb{R})$ that satisfies \eqref{KL_var}. Once such a $w^a$ is found, the Kirchhoff-Love Ansatz is used to construct 
\begin{equation}\label{kld}
u^a=w^a\textbf{e}_3-x_3\nabla w^a,
\end{equation}
a three-dimensional displacement field over $\Omega^r$ that is supposed to approximate the flexure part of the real displacement field $u^r$.

\subsection{Sequence $\{P_\varepsilon\}$.} To justify and validate the choice of the Kirchhoff-Love $P^a$,  
variational convergence has been used: $\Gamma-$convergence by Anzellotti {\it et al.}  \cite{ABP94} and by Bourquin {\it et al.}  \cite{BCGR92}; functional analysis methods by Ciarlet and coworkers  (for a comprehensive account, see \cite{Ci97}). The starting point of a $\Gamma-$convergence analysis is problem $P^r$ in its formulation \eqref{Pr_var}; the methods described in  \cite{Ci97} are based on the weak formulation  
\eqref{Pr_weak} of the same problem. We shall develop our considerations with reference to the former formulation, although our arguments could be easily rephrased so as to apply to convergence methods devised for the latter.

As stated in the Introduction, variational convergence studies the limit of a problem sequence $\{P_\eps\}$ indexed by a small parameter $\eps$ that is made to approach zero. Such a sequence is usually constructed in two steps: (i) a domain sequence  $\{\Omega_\eps\}$ is introduced, such that $\;\Omega_\eps\rightarrow\omega^r\,$ as $\,\eps\rightarrow 0$; (ii) for each domain $\Omega_\eps$, a functional $\mathcal{F}_\eps$ is defined, closely related to $\mathcal{F}^r$. Precisely, following \cite{ABP94, BCGR92}, in the first place one sets:
$$
\Omega_\eps=\omega^r\times \eps (-h^r,+h^r),\quad \eps\in (0,1],$$
so that the sequence of domains $\Omega_\eps$ is obtained by a homothetical rescaling of $\Omega^r$ with respect to thickness; secondly, one looks for an $\eps-$family of functionals $\mathcal{F}_\eps$. An easy way to have such a family would seem to take $\mathcal{F}_\eps$ to be $\mathcal{F}^r$ with   $\Omega^r$ replaced by $\Omega_\eps$:
\begin{equation}\label{unbfun}
v\mapsto \int_{\Omega_{\eps}}\Big(\frac 12\mathbb{C}^r [E(v)]\cdot E(v)- b^r\cdot v\Big) dx.
\end{equation}
However, this simplistic measure does not work, the reason being that keeping the loads independent of $\eps$  implies that the minimizers of functional \eqref{unbfun} become unbounded, and hence their sequence does not converge when thickness tends to null. This is the reason why, on a second attempt, the loads $b^r$ are replaced by a sequence of loads $b_\eps$ that are $\eps$-scaled  so as to keep the solutions $u_\eps$ bounded in a suitable norm when $\eps\rightarrow 0$. {In conclusion, the typical functional to be studied is:
\begin{equation}\label{efam}
\mathcal{F}_\eps(v):= \int_{\Omega_{\eps}}\Big(\frac 12\mathbb{C}^r [E(v)]\cdot E(v)- b_\eps\cdot v\Big) dx,\footnote{Note that, if $b_1=b^r$, then $\mathcal{F}_1=\mathcal{F}^r$.}
\end{equation}
and the related problem $P_\eps$ is:}
\begin{equation}\label{P_eps}
\textrm{find}\;\;u_\eps=\arg\!\!\!\!\!\!\!\!\!\!\min_{v\in H^1_D(\Omega_\eps;\mathbb{R}^3)}
\mathcal{F}_\eps(v).
\end{equation}

The above procedure to construct the sequence $\{P_\eps\}$  to be associated with a given problem $P^r$  was used in \cite{CiDe79} to achieve dimension reduction; within the framework of $\Gamma-$convergence, it was first employed in \cite{ABP91}. Later on, except for a few cases, the problem sequences considered in the literature on dimension reduction have been constructed as just described; in the following, we call them {\it classical sequences}. In the next subsection, we comment briefly on the meaning of solution convergence when such problem sequences are employed.

\subsection{Solution convergence.} 
Let 
$u_\eps$ be a solution of problem $P_\eps$ formulated in \eqref{P_eps}. Since $u_\eps$ is defined over domains which depend on $\eps$, the sentence:  ``$u_\eps$ converges to $u_0$, as $\eps\to 0$'' should be appropriately interpreted. We recall two of these interpretations here below.

\noindent - Anzellotti {\it et\ al.}  \cite{ABP94} define the operator 
$$
q_\eps:H^1(\Omega_\eps;\mathbb{R}^3)\to H^1(\omega^r;\mathbb{R}^3),\quad q_\eps(v)(x_1,x_2)=\frac{1}{2\eps {{h^r}}}\int_{-\eps{h^r}}^{\eps{h^r}}v(x_1,x_2,x_3)\,dx_3;
$$
$q_\eps$ associates to the field $v$ defined over $\Omega_\eps$ its \emph{fiber average} $q_\eps(v)$, a field defined over the fixed flat domain $\omega^r$.
Thus, 
\begin{equation}\label{conv_ABP}
\textrm{in \cite{ABP94}},\;\,u_\eps\to u_0 \,\mbox{ means that }\, q_\eps(u_\eps)\to u_0,
\end{equation}
in an appropriate topology, which needs not to be specified here; by this method the limit problem $P_0$ turns out to be posed on the two-dimensional domain $\omega^r$. 

\noindent -  Bourquin {\it et\ al.}  \cite{BCGR92} define a scaling map  
\begin{equation}\label{seps}
s_\eps:\Omega_1\to \Omega_\eps,\quad s_\eps(x_1,x_2,x_3):=(x_1,x_2,\eps x_3)\mag{,}
\end{equation} 
so that $u_\eps\circ s_\eps\in H^1(\Omega_1;\mathbb{R}^3)$.\footnote{Here, $\Omega_1=\Omega^r$. Note that
$$
s_\eps(\Omega_1)=\omega^r\times\varepsilon(-h^r,+h^r)=\Omega_\varepsilon.
$$
} Then,
\begin{equation}\label{conv_BCGR}
\textrm{in \cite{BCGR92}},\;\, u_\eps\to u_0 \,\mbox{ means that }\, u_\eps\circ s_\eps\to u_0,
\end{equation}
again, in a topology that it is not necessary to specify for our present discussion.

It seems to us important to realize that the limit displacement $u_0$ in \eqref{conv_ABP} is not the same as in \eqref{conv_BCGR}: in the latter case, the limit problem $P_0$, namely,
$$
\textrm{find}\;\;u_0=\arg\!\!\!\!\!\!\!\!\!\!\min_{v\in H^1_D(\Omega_1;\mathbb{R}^3)}
\mathcal{F}_0(v),
$$
is posed on $\Omega_1$ and not on $\omega^r$, as is the former case. However, the domain of the limit functional $\mathcal{F}_0$ turns out to be the space of Kirchhoff-Love displacements \eqref{kld}, and hence  problem $P_0$ can be
easily rewritten, via thickness integration, in terms of functions defined over $\omega^r$.

We also point out that, for the sake of keeping our discussion of solution convergence concise, we did not pause and detail  several technicalities, that anyway would not affect our arguments in any manner.
For instance,  neither we mentioned that Anzellotti {\it et\ al.}  \cite{ABP94} achieved their result by a $\Gamma-$asymptotic expansion nor that Bourquin {\it et \ al.}  \cite{BCGR92}, in addition to the coordinates of points in $\Omega_1$, did scale the components of $u_\eps$, again not all in the same way. The scaling of $u_\eps$ they adopted allowed these authors to deduce the desired result  by {taking only one $\Gamma$-limit, thereby avoiding the use of  $\Gamma$-asymptotic expansions.}

\section{Nonclassical problem sequences}\label{sec_ncs}

In the previous section, we have {exemplified} how, given a real problem $P^r$, a classical problem sequence $\{P_\varepsilon\}$ is constructed, which variationally converges to a limit problem $P_0$ {akin to $P^r$}.
In this section, we explain how $P_0$ is related to $P^r$ and show that the argument
outlined in Section \ref{sec_cs} can be extended to nonclassical sequences.

\subsection{Generalities.}\label{3.1} Let the given problem $P^r$ be {posed over} a three-dimensional domain $\Omega^r$ that is {\emph{thin}, in the sense that there are a one- or two-dimensional domain $\omega^r$ and a real number $\eps^r\ll 1$ such that $\textit{meas}(\Omega^r)\propto(\eps^r)^{2\,\textrm{or}\,1}\textit{meas}(\omega^r)$.}
We are interested in finding a problem $P_0$, {easier to solve than $P^r$,} whose solution $u_0$ is {guaranteed to be `close'} to the
solution $u^r$ of problem $P^r$.
\footnote{Here and henceforth, the notion of {solution `closeness'} is mathematically vague: only an error analysis can assess how good an approximation is. {As a matter of fact}, variational-convergence methods {can} identify 
a `good' {approximating} problem $P_0$, but they do not provide us with a careful estimate of the error implicit in replacing  $P^r$  with $P_0$. With these provisos, we shall use the words `close' and `small' freely.} {Here is} a two-step sequence of operations leading to obtain {such} a limit problem $P_0$ via variational convergence, for any given problem $P^r$.

\vskip 6pt
\noindent
\textsc{Step 1.}
Choose a sequence of domains $\Omega_\eps$ such that
\begin{itemize}
\item [(i)] $\Omega_\eps$ approaches $\Omega^r$ as $\eps$ goes to zero;
\item [(ii)] $\Omega_{\eps^r}=\Omega^r$.
\end{itemize}
\noindent
\textsc{Step 2.}
Choose a sequence of problems $P_\eps$ defined over $\Omega_\eps$, such that
\begin{itemize}
\item [(i)] {$\{P_\eps\}$} variationally converges;
\item [(ii)] $P_{\eps^r}=P^r$.
\end{itemize}

{We stress that the convergence requirements at points (i) of both steps concern} sequences of domains in Step 1, of problems in Step 2; and that points (ii) force the chosen {domain and problem} sequences to include, respectively,  the real domain $\Omega^r$ and the real problem $P^r$ for $\eps$ equal to the real $\eps^r$. {Moreover, an all-important fact is
that  sequence ${\{P_\varepsilon\}}$ may be quite \textit{artificial}, in that it must} have something in common with the
{real} problem of interest only for $\eps=\eps^r$.

{We now show that the two-step procedure we propose does yield} a problem $P_0$ close to $P^r$. {To begin with, according to} (i) in Step 2,  {$\{P_\eps\}$}  variationally converges to some limit problem; we denote such a limit problem by $P_0$, and its solution by $u_0$. {Because of }
\eqref{convar}, the sequence {$\{u_\eps\}$ of solutions to} problems $P_\eps$ converges to $u_0$; in particular,  $u_\eps$ is close to $u_0$  for $\eps$ small. But, since $\eps^r$ is small, we have by (2) that $u^r$, the solution of $P^r$, is equal to $u_{\eps^{r}}$, and hence that $u^{r}$ is close to $u_0$.
We have thus  `proved' that the solution of problem $P_0$ is close to the solution of 
problem $P^r$.

Note that, if $\eps$ is replaced with $\eps/\eps^r$, then {all classical} sequences {considered}  in Section \ref{sec_cs} {comply with  the procedure we laid down here above, which does nothing else than formalizing how a limit problem $P_0$  close to a real problem $P^r$ is found  by means of a classical sequence.}

Two features of our procedure call for attention:  

\noindent - \emph{it allows for constructing more than one problem sequence having the desired properties, each with its own limit problem.}

\noindent - \emph{it allows for constructing problem sequences having scarce physical meaning, if any.}  

The first feature should come to no surprise, and indeed be welcomed. Suppose two  sequences $\{P_\varepsilon\}$ and $\{\bar P_\varepsilon\}$ are constructed, with two different limit problems
$P_0$ and $\bar P_0$, both being close to $P^r$ in view of the above reasoning. 
Problem $P_0$ can be seen as an approximation of problem $P^r$; hence, necessarily, the solution $u_0$ of the former captures some of the features of the solution $u^r$ of the latter; on the other hand, by the same token, the solution $\bar u_0$ of  problem $\bar P_0$ may capture additional or different characters of $u^r$, or just the same characters with a different degree of accuracy. In fact, the choice of a sequence $\{P_\varepsilon\}$  decides which distinctive properties of $u^r$ are going to be preserved in the solution of the limit problem $P_0$, and in which detail. 

The second feature suggests a word of caution for potential users of our procedure. Here is why. Let 
the problems in sequence $\{\bar P_\eps\}$ be defined over a sequence of domains $\Omega_\eps$ such that $\Omega_{\eps^r}=\Omega^r$ and approaching the low-dimensional domain $\omega^r$ as $\eps$ goes to zero, and assume that $\{\bar P_\eps\}\rightarrow P^u$, with problem $P^u$ defined over  domain $\omega^r$; furthermore, let  the sequence $\{P_\eps\}$ be defined as follows:
$P_\eps=\bar P_\eps$ if $\eps\ne \eps^r$ and $P_{\eps^r}=P^r$. Clearly, this second sequence conforms to our procedure and converges to $P_0=P^u$.\footnote{We gratefully thank Fran\c{c}ois Murat for this remark.} Thus,  our recipe for construction of variationally convergent problem sequences should be regarded as a minimal collection of requirements: a priori, no complying sequence can be preferred  to any other just on mathematical grounds, it is for physical intuition to orient the selection. 

\subsection{Example 1.}
Let $P^r$ be the equilibrium problem of an \emph{isotropic} and linearly elastic plate-like body. Just as in Section \ref{2.1},  we let the body occupy a three-dimensional region $\Omega^r=\omega^r\times(-h^r,+h^r)$, whose cross section is $\omega^r$ and whose thickness is $2h^r$; moreover, for simplicity, we stipulate that the same mixed boundary conditions as in Section \ref{2.1} apply, and that the body is subject to the same type of loads. The only difference is that we now choose  the following well-known form of the stored-energy density:
\begin{equation}\label{isoen}
W^r(E)=\mu |E|^2+\frac \lambda 2 (\mbox{tr}\,E)^2,\quad \mu>0,\;\,3\lambda+2\mu>0,
\end{equation}
with $\lambda,\mu$ the Lam\'e moduli. Accordingly, we formulate the following real problem $P^r$:
\begin{equation}\label{Pr_variso}
\textrm{find}\;\;u^r=\,\arg\!\!\!\!\!\!\!\!\!\!\!\min_{u\in H^1_D(\Omega^r;\mathbb{R}^3)}\int_{\Omega^r}\Big(W(u)- b^r\cdot u\Big) dx,
\end{equation}
where
\begin{equation}\label{stoen}
\quad W(u):=W^r(E(u))=\mu |E(u)|^2+\frac \lambda 2 (\mbox{tr}\, E(u))^2.
\end{equation}

With a view toward applying our two-step procedure to associate with this one $P^r$ two different problem sequences, whose different limit problems have a long-standing status in structure engineering, we take care of Step 1 by letting
$$
\Omega_\eps=\omega^r\times \frac{\eps}{\eps^r}(-h^r,+h^r),\quad \eps\in(0,\eps^r].
$$
Next, we observe that \eqref{isoen} can be re-written as follows:
\begin{eqnarray*}
W^r(E)&=&\frac{2\mu+ \lambda}{ 2} (E_{11}+E_{22})^2-2\mu(E_{11}E_{22}-E_{12}^2)\\
&+&\frac{2\mu+ \lambda}{ 2} E_{33}^2+\lambda (E_{11}+E_{22}) E_{33}+2\mu (E_{13}^2+E_{23}^2).
\end{eqnarray*}
On adapting a line of reasoning inspired by this observation and first exploited in \cite{PPGT06, PPGT07}, we set:
\begin{eqnarray*}
\widehat W_\eps(E,u;\kappa)&=&\frac{2\mu+ \lambda}{ 2} (E_{11}+E_{22})^2-2\mu(E_{11}E_{22}-E_{12}^2)\\
&+&\frac{2\mu+ \lambda}{ 2} \left(1-\kappa+\kappa\left(\frac{\eps^r}{\eps}\right)^2\right) E_{33}^2
\\
&+&\lambda \left(1-\kappa+\kappa\left(\frac{\eps^r}{\eps}\right)\right)(E_{11}+E_{22}) E_{33}+2\mu(E_{13}^2+E_{23}^2)\\
&+&\kappa\left(\frac{\eps^r-\eps}{\eps}\right)^2\big((u_{1,33})^2+(u_{2,33})^2\big),\quad \kappa\geq 0
\end{eqnarray*}
(note that $\,\widehat W_{\eps^r}(E,u;\kappa)=W^r(E)$, whatever the value of  parameter $\kappa$). Furthermore, we define:
\[
W_\eps(u;\kappa)=\widehat W_\eps(E(u),u;\kappa),
\]
and we let $P_\eps(\kappa)$ be the typical representative of the following family of minimization problems:

\begin{equation}\label{Peps_variso}
\textrm{find}\;\;u_\eps=\,\arg\!\!\!\!\!\!\!\!\!\!\!\min_{u\in H^1_D(\Omega_\eps;\mathbb{R}^3)}
\frac{1}{\eps}\int_{\Omega_{\eps}}\Big(W_\eps(u;\kappa)- b_\eps\cdot u\Big) dx.
\end{equation}
Premultiplication by a constant never changes a functional's set of minimizers. In the present case, premultiplication by $1/\eps^{2\beta}$, with $\beta\in \Ro$, has the only effect of rescaling loads and displacements uniformly:
\[
\frac{1}{\eps^{2\beta}}\int_{\Omega_{\eps}}\!\!\!\Big(W_\eps(u;\kappa)- b_\eps\cdot u\Big) dx=\!\!
\int_{\Omega_{\eps}}\!\!\!\Big(W_\eps(\tilde u;\kappa)- \tilde b_\eps\cdot \tilde u\Big) dx,\quad \tilde u:={u}/{\eps^\beta},\;\,\tilde b_\eps={b_\eps}/{\eps^\beta};\footnote{
In the context of nonlinear elasticity, where the stored-energy density is not quadratic, the value of $\beta$ determines different `behavior regimes' (see \cite{FJM06}).}
\]
in particular, as shown in \cite{PPGT06, PPGT07}, choosing  $\beta=1/2$ implies that, for an appropriate load sequence $\{b_\eps\}$ and in an appropriate topology, which needs not be specified here,
\begin{equation}\label{defu0}
\frac{(u_\eps\circ s_\eps)_\alpha}{\eps}\to (u_0)_\alpha\quad\mbox{and}\quad (u_\eps\circ s_\eps)_3\to (u_0)_3
\end{equation}
 (in \eqref{defu0},  the map $s_\eps$ is as defined in \eqref{seps} and the index $\alpha=1,2$ is used to denote the in-plane components of the displacement vector).

It is not difficult to prove that the problem sequence $\{P_\eps(0)\}$ leads to the Kirchhoff-Love theory of \emph{unshearable} plates (cf. Anzellotti {\it et al.}  \cite{ABP94} and Bourquin {\it et al.}  \cite{BCGR92}), in which the limit displacement $u_0$, as defined by \eqref{defu0}, belongs to the space of Kirchoff-Love displacements
\begin{eqnarray*}
u_0\in \mathcal{KL}:=\{w^a\textbf{e}_3+\textbf{v}^a -x_3\nabla w^a\!\!\!&:&\!\!\!\textbf{v}^a\in H^1_D(\omega^r;\Ro^2), w^a\in H^2(\omega^r)\\
&&\!\!\!\mbox{and }w^a=w^a_{,n}=0 \,\mbox{ on }\partial_D\omega^r \}.
\end{eqnarray*}  
On the other hand, for $\kappa>0$, the problem sequence $\{P_\eps(\kappa)\}$ leads to a theory of \emph{shearable} plates  (cf. \cite{PPGT06, PPGT07}) in which the limit displacement $u_0$ belongs to the space of Reissner-Mindlin displacements, i.e.,
\begin{eqnarray*}
u_0\in \mathcal{RM}:=\{w^a\textbf{e}_3+\textbf{v}^a +x_3\boldsymbol{\varphi}^a\!\!\!&:&\!\!\!\textbf{v}^a,\boldsymbol{\varphi}^a\in H^1_D(\omega^r;\Ro^2),\; w^a\in H^1_D(\omega^r)\}.
\end{eqnarray*}  

\vskip 10pt
\subsection{Example 2.} In the previous example, two problem sequences were constructed by associating with the same domain sequence two different sequences of energy densities. We now consider two domain sequences and associate  the same sequence of energy densities with both. 

\begin{figure} [h]
\centering 
\def\svgwidth{\columnwidth} 
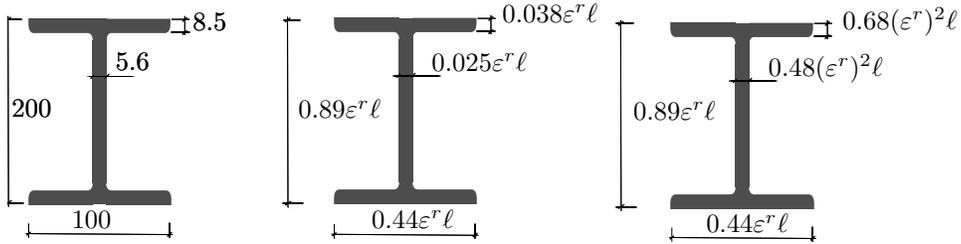
\caption{The IPE200 ``double-T'' cross section (dimensions in \emph{mm}).}
\label{fig2}
\end{figure}

Let $\Omega^r$ be a space region in the form of a  right cylinder, occupied by a linearly elastic beam-like body having a IPE200 ``double-T'' cross section (see Fig. \ref{fig2} $(i)$) 
and length $\ell=4 m$. No doubt, region $\Omega^r$ is thin, in that the ratio of its cross-section diameter to its length is  $\eps^r=\sqrt{100^2+200^2}/4000\approx0.056$. Another beam-like body made of the same material, of the same length but with $100\times200$ rectangular cross section, must be considered equally thin according to such a thinness notion; in fact, the response to bending loads of the two bodies would not differ much. Not so for their response to twisting loads, that would turn out to be sensitive to the cross-section shape, the former body exhibiting a larger rotation. Now, one can choose whether or not to come up with a variational limit under form of a beam theory that incorporates the cross-section shape effects we described, and others. 

In case there is no need for a theory capable of detailed predictions, it is sufficient to note that in terms of $\eps^r$ the cross-section width and height read as in Fig. \ref{fig2} $(ii)$; to let $\omega_\eps$ be the sequence of two-dimensional domains obtained by replacing $\eps^r$ 
by $\eps$; and to let $\Omega_\eps=\omega_\eps\times (0,\ell)$. Then, a classical problem sequence of type  \eqref{P_eps}  leads to the Bernoulli-Navier theory of beams (see Percivale \cite{Pe99}).

Otherwise, a subtler domain rescaling is in order. For instance, one writes thickness of web and wings of the ``double-T'' cross section under study in terms  of $(\eps^r)^2$ instead of $\eps^r$, as shown in Fig. \ref{fig2} $(iii)$ (note that, in so doing, all the
scaling coefficients become ``comparable"). 
Moreover, for $\bar\omega_\eps$ the sequence of two-dimensional domains obtained by replacing $\eps^r$ in Fig. \ref{fig2} $(iii)$
by $\eps$, one lets $\bar\Omega_\eps=\bar\omega_\eps\times (0,\ell)$ and denotes by $\bar P_\eps$ the relative classical problem sequence of type \eqref{P_eps}. This problem sequence of leads to the Vlassov beam theory, 
(see Freddi \emph{et al.} \cite{FMP04,FMP07}).

Just as in Example 1, by considering two different problem sequences we ended up with two quite different model problems: on the mechanical side, cross sections remain plane in Bernoulli-Navier's theory, while Vlassov's theory allows for their deformation; on the mathematical side, the domain of definition of Bernoulli-Navier's and Vlassov's energies are different, because the twist angle, as a function of the axial coordinate,  is required to be once differentiable  in the former theory, twice in the latter.

\subsection{Further comments.}\label{3.4} So far, we have exemplified the use for dimension-reduction problems of the minimal recipe provided in Section \ref{3.1}. As mentioned in the Introduction, the same recipe works also for other problem classes; we sketch how it does in the case of \emph{periodic homogenization}. 

Let $\Omega^r$ be the reference configuration of a composite material body, i.e., an inhomogeneous body, whose material properties are periodic in space; moreover, let $\ell^r$ be a characteristic length of the periodicity cell -- its diameter, say -- and let $\eps^r$ be the ratio between $\ell^r$ and the diameter of $\Omega^r$, so that, typically, $\eps^r\ll1$. Then, for $P^r$ any real problem defined over the space region $\Omega^r$, we can apply our minimal recipe to find the corresponding homogenized problem $P_0$, by taking   $\Omega_\eps\equiv \Omega^r$ in the domain sequence at Step 1  and by choosing for $\{P_\eps\}$ any problem sequence satisfying the requirements listed under Step 2.

\vskip 6pt
 
The approach of Braides and Truskinovsky \cite{BT08} slightly intersects ours, in that their starting point is a given problem sequence  and its $\Gamma$-limit (essentially, what we here call a classical sequence $\{P_\eps\}$ with limit problem $P_0$). A main goal of theirs, among others, is to set up a so-called \emph{$\Gamma$-development} of the given problem sequence, that is, to say it simply, a procedure that delivers a representation of  $\{P_\eps\}$  up to some prescribed order $\eps^\alpha$, in the form $\{P_\eps=P_0+\eps^\alpha P^{(\alpha)}+ o(\eps^\alpha)\}$, whence, hopefully, a better variational approximation of the minimization problem $P_\eps$ than the zero-order problem $P_0$ would ensue.

\section{Final Remarks}
We have shown how one given real problem can be associated with various problem sequences, whose variational-limit problems are akin to different low-dimensional models from the theory of elastic structures. Rather than scary, this freedom should be regarded as potentially beneficial, because it may help fixing some unphysical
results. 

This is the case, for instance, in the deduction by dimension reduction of the motion equations for a linearly elastic plate, when the use of the classical-sequence approach leads to an evolution equation only for out-of-plane motions, whereas the quasi-static equations for in-plane motions feature no inertia terms.
The reason for this is essentially intrinsic to the following well-known weak convergences {(cf. \cite{Ci97, Pa06})}:

$\displaystyle \frac{u_\alpha^\eps}{\eps}\rightharpoonup u_\alpha$ for
 in-plane displacements, 
$\displaystyle {u_3^\eps}\rightharpoonup u_3$ for  out-of-plane displacements,
which imply that, whatever the test function $\psi$, the inertial working
$$
\int (-\rho \stackrel{..}u^\eps)\cdot \psi\,dx=-\int \rho \Big({\ddot u}^\eps_3 \psi_3+\eps \frac{\stackrel{..}u^\eps_\alpha}{\eps} \psi_\alpha\Big)dx
$$
converges to 
$$
-\int \rho \stackrel{..}u_3 \psi_3\,dx.
$$
The fact that this limit term contains no in-plane inertial contribution conflicts with experience,
 because in-plane waves of measurable velocity do propagate in a plate. A remedy consists in considering problem sequences different from classical and yet compliant with our proposed recipe, e.g., a sequence where the inertial working is: 
$$
-\int \rho \Big({\ddot u}^\eps_3 \psi_3+\left(\frac{\eps^r}{\eps}\right){\ddot u}^\eps_\alpha \psi_\alpha\Big)dx=
-\int \rho \Big({\ddot u}^\eps_3 \psi_3+\eps^r \frac{{\ddot u}^\eps_\alpha}{\eps} \psi_\alpha\Big)dx.
$$

A final question arises: which of the several sequences we may associate with a given real problem is the best one? As every other `natural' question, this is ill-posed, unless an optimality criterion is stipulated. Such a stipulation presumes that those features of the real problem that one especially wishes to approximate are chosen, be they the displacement field, the stress field, or other; and that an error measure is selected, in terms of an appropriate norm. Then, the best sequence is the one that delivers a limit problem whose solution is the closest in norm to the real solution.


\begin{thebibliography}{10}

\bibitem{ABP91}
E.~Acerbi, G.~Buttazzo, and D.~Percivale.
\newblock A variational definition of the strain energy for an elastic string.
\newblock {\em J. Elasticity}, 25(2):137--148, 1991.

\bibitem{ABP94}
G.~Anzellotti, S.~Baldo, and D.~Percivale.
\newblock Dimension reduction in variational problems, asymptotic development
  in {$\Gamma$}-convergence and thin structures in elasticity.
\newblock {\em Asymptotic Anal.}, 9(1):61--100, 1994.

\bibitem{BCGR92}
F.~Bourquin, P.~G. Ciarlet, G.~Geymonat, and A.~Raoult.
\newblock {$\Gamma$}-convergence et analyse asymptotique des plaques minces.
\newblock {\em C. R. Acad. Sci. Paris S\'er. I Math.}, 315(9):1017--1024, 1992.

\bibitem{BDMG99}
A.~Braides, G.~Dal~Maso, and A.~Garroni.
\newblock Variational formulation of softening phenomena in fracture mechanics:
  the one-dimensional case.
\newblock {\em Arch. Ration. Mech. Anal.}, 146(1):23--58, 1999.

\bibitem{BrGe04}
A.~Braides and M.~S. Gelli.
\newblock The passage from discrete to continuous variational problems: a
  nonlinear homogenization process.
\newblock In {\em Nonlinear homogenization and its applications to composites,
  polycrystals and smart materials}, volume 170 of {\em NATO Sci. Ser. II Math.
  Phys. Chem.}, pages 45--63. Kluwer Acad. Publ., Dordrecht, 2004.

\bibitem{BT08}
A.~Braides and L.~Truskinovsky.
\newblock Asymptotic expansions by {$\Gamma$}-convergence.
\newblock {\em Contin. Mech. Thermodyn.}, 20(1):21--62, 2008.

\bibitem{Ci97}
P.~G. Ciarlet.
\newblock {\em Mathematical elasticity. {V}ol. {II}}, volume~27 of {\em Studies
  in Mathematics and its Applications}.
\newblock North-Holland Publishing Co., Amsterdam, 1997.
\newblock Theory of plates.

\bibitem{CiDe79}
P.~G. Ciarlet and P.~Destuynder.
\newblock A justification of the two-dimensional linear plate model.
\newblock {\em J. M\'ecanique}, 18(2):315--344, 1979.

\bibitem{DGFr75}
E.~De~Giorgi and T.~Franzoni.
\newblock Su un tipo di convergenza variazionale.
\newblock {\em Atti Accad. Naz. Lincei Rend. Cl. Sci. Fis. Mat. Natur. (8)},
  58(6):842--850, 1975.

\bibitem{DGSp73}
E.~De~Giorgi and S.~Spagnolo.
\newblock Sulla convergenza degli integrali dell'energia per operatori
  ellittici del secondo ordine.
\newblock {\em Boll. Un. Mat. Ital. (4)}, 8:391--411, 1973.

\bibitem{FMP04}
L.~Freddi, A.~Morassi, and R.~Paroni.
\newblock Thin-walled beams: the case of the rectangular cross-section.
\newblock {\em J. Elasticity}, 76(1):45--66 (2005), 2004.

\bibitem{FMP07}
L.~Freddi, A.~Morassi, and R.~Paroni.
\newblock Thin-walled beams: a derivation of {V}lassov theory via
  {$\Gamma$}-convergence.
\newblock {\em J. Elasticity}, 86(3):263--296, 2007.

\bibitem{FJM02}
G.~Friesecke, R.~D. James, and S.~M{\"u}ller.
\newblock A theorem on geometric rigidity and the derivation of nonlinear plate
  theory from three-dimensional elasticity.
\newblock {\em Comm. Pure Appl. Math.}, 55(11):1461--1506, 2002.

\bibitem{FJM06}
G.~Friesecke, R.~D. James, and S.~M{\"u}ller.
\newblock A hierarchy of plate models derived from nonlinear elasticity by
  gamma-convergence.
\newblock {\em Arch. Ration. Mech. Anal.}, 180(2):183--236, 2006.

\bibitem{LDRa95}
H.~Le~Dret and A.~Raoult.
\newblock The nonlinear membrane model as variational limit of nonlinear
  three-dimensional elasticity.
\newblock {\em J. Math. Pures Appl. (9)}, 74(6):549--578, 1995.

\bibitem{MPG}
B.~Miara and P.~Podio-Guidugli.
\newblock Deduction by scaling: a unified approach to classic plate and rod
  theories.
\newblock {\em Asymptotic Analysis}, 51(2):113--131, 2007.

\bibitem{MoMo77}
L.~Modica and S.~Mortola.
\newblock Un esempio di {$\Gamma ^{-}$}-convergenza.
\newblock {\em Boll. Un. Mat. Ital. B (5)}, 14(1):285--299, 1977.

\bibitem{MuTa77}
F.~Murat.
\newblock H-convergence.
\newblock {\em S\'eminaire dÕanalyse fonctionnelle et num\'erique, Universit\'e
  dÕAlger, 1977-78. {\rm English translation Murat F. and Tartar L.,
  H-convergence, Topics in the mathematical modelling of composite materials,
  21Ð43, Progr. Nonlinear Differential Equations Appl., 31, Birkhauser, Boston,
  MA, 1997.}}, 8:391--411, 1973.

\bibitem{Pa06}
R.~Paroni.
\newblock The equations of motion of a plate with residual stress.
\newblock {\em Meccanica}, 41(1):1--21, 2006.

\bibitem{PPGT06}
R.~Paroni, P.~Podio-Guidugli, and G.~Tomassetti.
\newblock The {R}eissner-{M}indlin plate theory via {$\Gamma$}-convergence.
\newblock {\em C. R. Math. Acad. Sci. Paris}, 343(6):437--440, 2006.

\bibitem{PPGT07}
R.~Paroni, P.~Podio-Guidugli, and G.~Tomassetti.
\newblock A justification of the {R}eissner-{M}indline plate theory through
  variational convergence.
\newblock {\em Anal. Appl. (Singap.)}, 5(2):165--182, 2007.

\bibitem{Pe99}
D.~Percivale.
\newblock Thin elastic beams: the variational approach to {S}t. {V}enant's
  problem.
\newblock {\em Asymptot. Anal.}, 20(1):39--59, 1999.

\bibitem{PePG09}
D.~Percivale and P.~Podio-Guidugli.
\newblock A general linear theory of elastic plates and its variational
  validation.
\newblock {\em Boll. Unione Mat. Ital. (9)}, 2(2):321--341, 2009.

\bibitem{PGJEla}
P.~Podio-Guidugli.
\newblock An exact derivation of the thin plate equation.
\newblock {\em J. Elasticity}, 22:121--133, 1989.

\bibitem{PGCism}
P.~Podio-Guidugli.
\newblock Concepts in the mechanics of thin structures.
\newblock CISM Volume 503, A. Morassi and R. Paroni (Eds.):77--110, 2008.

\bibitem{Sp68}
S.~Spagnolo.
\newblock Sulla convergenza di soluzioni di equazioni paraboliche ed
  ellittiche.
\newblock {\em Ann. Scuola Norm. Sup. Pisa (3) 22 (1968), 571-597; errata,
  ibid. (3)}, 22:673, 1968.

\bibitem{Ta09}
L.~Tartar.
\newblock {\em The general theory of homogenization}, volume~7 of {\em Lecture
  Notes of the Unione Matematica Italiana}.
\newblock Springer-Verlag, Berlin, 2009.
\newblock A personalized introduction.

\end{thebibliography}
%
\bibliographystyle{plain}

\end{document}